\documentclass[aps,prd,preprintnumbers,superscriptaddress,nofootinbib]{revtex4}
\usepackage[dvips]{graphicx}
\usepackage{bm,latexsym,amsmath,amssymb,amsfonts,mathrsfs}
\usepackage{color}
\input{colordvi.tex}

\newcommand{\bk}{{\bf k}}
\newcommand{\bp}{{\bf p}}
\newcommand{\bq}{{\bf q}}

\newcommand{\rL}{{\rm L}}
\newcommand{\rM}{{\rm m}}
\begin{document}

\title{Scale-dependent bias with higher order primordial non-Gaussianity:\\
 Use of the Integrated Perturbation Theory
}

\author{Shuichiro~Yokoyama}
\email[Email: ]{shu"at"icrr.u-tokyo.ac.jp}
\affiliation{Institute for Cosmic Ray Research, The University of Tokyo,
Kashiwa, Chiba, 277-8582, Japan}

\author{Takahiko~Matsubara}
\email[Email: ]{taka"at"kmi.nagoya-u.ac.jp}
\affiliation{Kobayashi-Maskawa Institute for the Origin of Particles and the Universe, Nagoya University, Chikusa, Nagoya, 464-8602, Japan}
\affiliation{Department of Physics, Nagoya University, Chikusa, Nagoya, 464-8602, Japan}

\begin{abstract}

We analytically derive a more accurate formula for the power spectrum of the biased objects with the primordial non-Gaussianity
parameterized not only by the non-linearity parameter $f_{\rm NL}$,
but also by $g_{\rm NL}$ and $\tau_{\rm NL}$ which characterize the trispectrum of the
primordial curvature perturbations.
We adopt the integrated perturbation theory which was constructed in Matsubara (2011)
\cite{Matsubara:2011ck}. 
We discuss 
an inequality between $f_{\rm NL}$ and $\tau_{\rm NL}$
in the context of the scale-dependent bias, by introducing a stochasticity parameter.
We also mention higher order loop corrections into
the scale-dependency of the bias parameter.
\end{abstract}

\preprint{ICRR-Report-631-2012-20}
\maketitle

\section{Introduction}

Primordial non-Gaussianities have been widely well-studied 
as new probes of the mechanism of generating primordial curvature perturbation, e.g., inflationary physics.
The observation of the cosmic microwave background (CMB) anisotropies
is one of the best ways to hunt for the primordial non-Gaussianities 
through the measurement of the bi- and tri-spectra of the temperature anisotropies.
For example, a current best limit for the primordial non-Gaussianity
is obtained as $-10 <f_{\rm NL} < 74$ at $95\%$ confidence level (CL) \cite{Komatsu:2010fb}
with $f_{\rm NL}$
being a local-type non-linearity parameter,
which characterizes the bispectrum of the primordial curvature perturbations.
The standard single field inflation model predicts that $f_{\rm NL}$
is the order of $10^{-2}$ and hence if we would detect that $f_{\rm NL} > O(1)$
it implies that our universe has not experienced the standard inflationary stage with a
single scalar field.

Thanks to the progress of the cosmological observations, recently,
the large scale structure (LSS) of the universe will also bring us
the fruitful information about the primordial fluctuations.
While
in the observation of LSS
the understanding of the gravitational non-linear evolution of the density fluctuations
is more important on smaller scales, 
the primordial non-Gaussianities affect the clustering of the biased objects on large scales.
It has been found that the local-type primordial non-Gaussianity, $f_{\rm NL}$,
produces significant scale-dependence on the bias parameter, $\Delta b \propto 1/k^2$.
The constraints on the primordial non-Gaussianities obtained from the observations
of such scale-dependence of the bias are comparable to those from the CMB observations,
e.g., $-31 < f_{\rm NL}  < 70 $ at $95\%$ CL in Ref. \cite{Slosar:2008hx}.

However,
in order to make ready for the upcoming precise observational cosmology,
on the theoretical side,
more deeply understanding of the effect of the primordial non-Gaussianities
should be necessary.
There are a lot of literatures about the precise formula for the scale-dependent bias
both analytically and numerically and also the analysis of the higher order effects.
Recently, Matsubara (2012)~\cite{Matsubara:2012nc} has shown an accurate formula of the scale-
dependent bias with primordial non-Gaussianity parametrized by $f_{\rm NL}$, by making use of the integrated Perturbation Theory
(iPT)~\cite{Matsubara:2011ck}.
By adopting this formula, it is not necessary to use the high peak approximation
and also the peak-background split picture, both of which have been considered as useful tools to
derive the bias parameter in the case with the primordial non-Gaussianity.

Together with deriving the precise formula of the bias parameter for $f_{\rm NL}$ case
(see, e.g., \cite{Desjacques:2011jb,D'Aloisio:2012hr}),
it should be important to investigate the effects of the higher order primordial non-Gaussianities,
that is, the trispectrum of the primordial curvature perturbations.
The non-linearity parameters denoted by $g_{\rm NL}$ and $\tau_{\rm NL}$
were introduced to characterize the amplitude of the primordial trispectrum.
There are also lots of literatures which study these non-linearity parameters theoretically and observationally.
 One of the interesting theoretical issue about these higher order non-Gaussianities
 is an inequality between $f_{\rm NL}$ and $\tau_{\rm NL}$, so-called, Suyama-Yamaguchi inequality
 \cite{Suyama:2007bg,Suyama:2010uj,Sugiyama:2011jt}.
 If the equality $\tau_{\rm NL} = 36 f_{\rm NL}^2 / 25$ is justified,
 the primordial curvature perturbations should be sourced from a single scalar field.
The case with $\tau_{\rm NL} > 36 f_{\rm NL}^2 / 25$ indicates that
the primordial curvature perturbations might be sourced from multi-scalar fields.

There have been several works about
the effects of these higher order parameters $g_{\rm NL}$ and $\tau_{\rm NL}$
on the scale-dependency of the bias parameter.
In Ref. \cite{Desjacques:2009jb},
the authors derive a scale-dependent bias in the case with non-zero $g_{\rm NL}$
and find that $g_{\rm NL}$ gives the same scale-dependency as the case with $f_{\rm NL}$.
They also give a constraint for $g_{\rm NL}$ by using the same data set as in Ref. \cite{Slosar:2008hx}.
For $\tau_{\rm NL}$, although there are not any available constraints, 
Refs. \cite{Gong:2011gx,Yokoyama:2011qr} analytically derive the formulae
of the scale-dependent bias with $\tau_{\rm NL}$ (Refs. \cite{Tseliakhovich:2010kf,Biagetti:2012xy} as related works).
Numerical analysis for the effects of such higher order non-Gaussianities
on LSS has been shown in Refs. \cite{Smith:2010gx,LoVerde:2011iz,Smith:2011ub,Nishimichi:2012da}.
However, comparing with $f_{\rm NL}$ case, the precise formulae for $g_{\rm NL}$
and $\tau_{\rm NL}$ cases might be less understood.
 
In this paper, we extend the analysis of the scale-dependent bias in Ref. \cite{Matsubara:2012nc}
to the case with higher order non-Gaussianities  characterized by
the non-linearity parameters $\tau_{\rm NL}$ and $g_{\rm NL}$, by making use of 
iPT.
In terms of iPT, the effects of $g_{\rm NL}$ and $\tau_{\rm NL}$
appear in two-loop order contributions and hence we also investigate the other contributions from
the higher order loops.
 
This paper is organized as follows. In the next section, we derive a power spectrum
of the biased objects with including the higher order primordial non-Gaussianities,
by making use of iPT.
In section \ref{sec:nGeffect},
we focus on the leading order contributions of $g_{\rm NL}$ and $\tau_{\rm NL}$,
and discuss the scale- and redshift-dependence of the bias parameters.
We introduce the stochasticity parameter
in order to investigate the inequality between $f_{\rm NL}$ and $\tau_{\rm NL}$.
In section \ref{sec:correction},
we discuss the higher order effects which are linearly proportional to the non-linearity parameters.
Section \ref{sec:summary} is devoted to summary and discussions.
We plot the figures of this paper with adopting the best fit cosmological parameters taken from
WMAP 7-year data \cite{Komatsu:2010fb}.

\section{Power spectrum of the biased objects with the primordial non-Gaussianities}

Here, we focus on the local-type primordial non-Gaussianity because
the several literatures have shown that other types of non-Gaussianities such as equilateral-,
orthogonal- and folded-types are not so effective on the structure formation of the Universe,
compared with the local type.

\subsection{Local-type non-Gaussianity}

Let us introduce the local-type non-linearity parameters.
The primordial non-Gaussianities of the curvature perturbations, $\Phi$, 
are usually characterized by the higher order spectra as
\begin{eqnarray}
&&\langle \Phi(\bk) \Phi(\bk')\rangle = (2\pi)^3 \delta^{(3)}(\bk + \bk') P_\Phi(k), \cr\cr
&&\langle \Phi(\bk_1) \Phi(\bk_2) \Phi(\bk_3) \rangle  
= (2\pi)^3 \delta^{(3)}(\bk_1 + \bk_2 + \bk_3) B_\Phi(\bk_1,\bk_2,\bk_3), \cr\cr
&&\langle \Phi(\bk_1) \Phi(\bk_2) \Phi(\bk_3) \Phi(\bk_4)\rangle_c  
= (2\pi)^3 \delta^{(3)}(\bk_1 + \bk_2 + \bk_3 + \bk_4) T_\Phi(\bk_1,\bk_2,\bk_3,\bk_4),
\end{eqnarray}
where a index, $c$, denotes a connected part of the correlation functions.
For the local-type case,
the higher order spectra are parameterized as
\begin{eqnarray}
B_\Phi(k_1,k_2,k_3) &=& 2 f_{\rm NL} \left[ P_\Phi(k_1)P_\Phi(k_2)+P_\Phi(k_2)P_\Phi(k_3)+P_\Phi(k_3)P_\Phi(k_1) \right], \cr\cr
T_\Phi(k_1,k_2,k_3,k_4) &=&
6 g_{\rm NL} \left[ P_\Phi(k_1)P_\Phi(k_2)P_\Phi(k_3) + ~3~{\rm perms.}
\right] \cr\cr
&& + {25 \over 9}\tau_{\rm NL} \left[  P_\Phi(k_1)P_\Phi(k_2)P_\Phi(|\bk_1+\bk_3|)
+ ~11~{\rm perms}. \right].
\end{eqnarray}
In the above expression, we call scale-independent parameters, $f_{\rm NL}$, $g_{\rm NL}$ and $\tau_{\rm NL}$, as the local-type non-linearity parameters.
As the simplest local-type model,
we consider that
 the primordial fluctuations are expanded as
\begin{eqnarray}
\Phi ({\bf x}) = \Phi_{\rm G} + f_{\rm NL} 
\left( \Phi_{\rm G}({\bf x})^2 -  \langle \Phi_{\rm G}({\bf x})^2\rangle \right)
+ g_{\rm NL} \Phi_{\rm G}({\bf x})^3 + \cdots,
\label{eq:single}
\end{eqnarray}
where $\Phi_{\rm G}$ denotes pure Gaussian fluctuations.
We have a special inequality for the local-type non-Gaussianity which is given by \cite{Suyama:2007bg,Suyama:2010uj,Sugiyama:2011jt}
\begin{eqnarray}
\tau_{\rm NL} \ge {36 \over 25} f_{\rm NL}^2.
\end{eqnarray}
For the case where the primordial curvature fluctuations are sourced from the quantum
fluctuations of a single scalar field and $\Phi$ is given by Eq.~(\ref{eq:single}),
we can realize the equality of the above relation.
If there exist the multiple scalar fields in the early Universe,
which are seeds of the primordial curvature fluctuations,
we have $\tau_{\rm NL} > 36/25 f_{\rm NL}^2$.
For example, in the case with
\begin{eqnarray}
\Phi = \phi_{\rm G}({\bf x}) + \psi_{\rm G}({\bf x})
 + f_{\rm NL}^{\phi} \left( \phi_{\rm G}({\bf x})^2 - \langle \phi_{\rm G}({\bf x})^2 \rangle \right),
\end{eqnarray}
we have
\begin{eqnarray}
f_{\rm NL} = { f_{\rm NL}^\phi \over (1+\xi)^2},~\tau_{\rm NL} = {36 \over 25}
 {\left({f_{\rm NL}^\phi}\right)^2 \over (1+\xi)^3},
 \end{eqnarray}
 and
\begin{eqnarray}
\tau_{\rm NL} &=& {36 \over 25} (1+\xi) f_{\rm NL}^2 \cr\cr
&>& {36 \over 25} f_{\rm NL}^2,~({\rm for}~\xi>0)
\end{eqnarray}
where we have used $\langle \phi \psi \rangle = 0$
and $\xi \equiv P_\psi / P_\phi$ with $P_i (i = \phi,\psi)$ being the power spectrum
of each component.
Hence, because of this fact, it is important to investigate the relation between
$\tau_{\rm NL}$ and $f_{\rm NL}$
for studying the origin of the primordial fluctuations.

\subsection{The power spectrum of the biased objects in integrated perturbation theory}

Let us derive a formula of the power spectrum of the biased objects
with the primordial non-Gaussianity by making use of iPT.
Following Refs. \cite{Matsubara:2011ck,Matsubara:2012nc},
we introduce 
multipoint propagators $\Gamma_X^{(n)}$ of the biased objects $X$ in "Eulerian" space which is given by
\begin{eqnarray}
\langle
{\delta^n \delta_X(\bk) \over \delta \delta_\rL(\bk_1) \cdots \delta \delta_\rL(\bk_n)}
\rangle
= (2\pi)^{3-3n} \delta^{(3)}(\bk_1+\cdots + \bk_n - \bk) \Gamma_X^{(n)}(\bk_1, \cdots , \bk_n),
\label{eq:multi}
\end{eqnarray}
where
$\delta_\rL(\bk)$ and $\delta_X(\bk)$ are respectively the Fourier transforms of the linear density field
and the number density field of the biased objects in "Eulerian" space.
The linear density field $\delta_\rL(\bk)$ is given by
\begin{eqnarray}
\delta_\rL (\bk) = {\cal M}(k) \Phi(\bk).
\end{eqnarray}
Here, the proportional factor 
${\cal M}(k)$ is given by
\begin{eqnarray}
{\cal M}(k) = {2 \over 3} {D(z) \over D(z_*)(1+z_*)} {k^2 T(k) \over H_0^2 \Omega_{\rM 0}},
\end{eqnarray}
where $T(k)$, $D(z)$, $H_0$ and $\Omega_{\rM 0}$ are respectively the transfer function, 
the linear growth factor,
the Hubble's constant and 
the matter density parameter. 
Here $z_*$ denotes an arbitrary redshift in the matter-dominated era.
 
The power-, bi- and tri- spectra of $\delta_\rL(\bk)$ are respectively given by
\begin{eqnarray}
P_\rL (k) &=& {\cal M}(k)^2 P_\Phi(k), \cr\cr
B_\rL (\bk_1, \bk_2, \bk_3) &=& {\cal M}(k_1){\cal M}(k_2){\cal M}(k_3)
B_\Phi (\bk_1, \bk_2, \bk_3), \cr\cr
T_\rL (\bk_1,\bk_2,\bk_3,\bk_4) &=&
{\cal M}(k_1){\cal M}(k_2){\cal M}(k_3){\cal M}(k_4)
T_\Phi (\bk_1, \bk_2, \bk_3,\bk_4).
\end{eqnarray}
Up to the leading order of the non-linearity parameters 
$f_{\rm NL}$, $g_{\rm NL}$ and $\tau_{\rm NL}$,
in terms of the multipoint propagators,
the power spectrum of the biased objects 
is given by
\begin{eqnarray}
P_X(k) &=&
\left[ \Gamma_X^{(1)}(\bk) \right]^2P_\rL(k) +
\Gamma_X^{(1)}(\bk) \int {d^3 p \over (2 \pi)^3}
\Gamma_X^{(2)}(\bp, \bk - \bp) B_\rL (\bk, -\bp, -\bk + \bp ) \cr\cr
&&
+ {1 \over 3} \Gamma_X^{(1)}(\bk)
\int {d^3 p_1 d^3 p_2 \over (2 \pi)^6}
\Gamma_X^{(3)} (\bp_1, \bp_2, \bk - \bp_1 - \bp_2)
T_\rL(\bk, - \bp_1, -\bp_2, - \bk + \bp_1 + \bp_2) \cr\cr
&& 
+ {1 \over 4}
\int {d^3 p_1 d^3 p_2 \over (2 \pi)^6}
\Gamma_X^{(2)} (\bp_1, \bk - \bp_1) 
\Gamma_X^{(2)}(-\bp_2, - \bk + \bp_2)
T_\rL(\bp_1,\bk-\bp_1,-\bp_2, -\bk+\bp_2)
 .
 \label{eq:power}
\end{eqnarray}
While the first line in the above equation has been already obtained
and investigated in details
 in Ref. \cite{Matsubara:2012nc},
 here, we look more deeply into
  the second and third lines.

The multipoint propagators include the non-linear evolution of the matter density field
and also the bias functions.
In order to derive a more useful expression for the power spectrum,
let us 
introduce renormalized bias function in "Lagrangian" space following Ref. \cite{Matsubara:2012nc},
which is defined as
\begin{eqnarray}
c_n^\rL(\bk_1,\bk_2, \cdots , \bk_n)
=
(2\pi)^{3n}
\int {d^3 p \over (2\pi)^3} \langle
{\delta^n \delta_X^\rL (\bp) \over \delta \delta_\rL(\bk_1) \cdots \delta \delta_\rL(\bk_n)}
\rangle,
\label{eq:renbias}
\end{eqnarray}
where $\delta_X^\rL$ is the number density field of the biased objects in "Lagrangian" space.
The multipoint propagators of the biased objects on large scales ($\bk_1+\bk_2+\cdots + \bk_n \to 0$),
where the non-linear evolution of the matter density field is negligible,
are written in terms of $c_n^\rL$ as
\begin{eqnarray}
\Gamma_X^{(1)}(\bk) & \to& 1 + c_{1}^\rL (\bk), \cr\cr
\Gamma_X^{(n)}(\bk_1,\bk_2, \cdots, \bk_n) &\to & c_n^\rL (\bk_1,\bk_2, \cdots, \bk_n)~{\rm for}~n \ge 2.
\label{eq:multiren}
\end{eqnarray}
From Eq. (\ref{eq:renbias}),
the renormalized bias function is calculated
once the relation between the linear density field and the number density field of the
biased objects in "Lagrangian" space is determined.
Based on the Press-Schechter (PS) picture, 
the mass function is determined by the probability that
the value of the "smoothed" linear density field over mass scale $M$
exceeds a critical value $\delta_c$, denoted by $P(M,\delta_c)$.
The smoothed linear density field over mass scale $M$ is given by
\begin{eqnarray}
\delta_M = \int {d^3 k \over (2\pi)^3} e^{i\bk\cdot {\bf x}} W(kR) \delta_\rL(\bk),
\end{eqnarray}
where $R$ is a comoving scale corresponding to the mass scale $M$
and $W(kR)$ is a usual top-hat window function.
Following Ref. \cite{Matsubara:2012nc}, the renormalized bias function can be written as
\begin{eqnarray}
c_n^\rL (\bk_1, \cdot, \bk_n) ={(-1)^n 
{\partial \over \partial M}\left[ {\partial^n P(M,\delta_c) \over \partial \delta_c^n}  W(k_1 R) \cdots W(k_n R) \right] \over \partial P(M,\delta_c)/ \partial M}.
\end{eqnarray}
 Assuming a universal mass function, we have
 \begin{eqnarray}
 {\partial P(M,\delta_c) \over \partial M} &=& {f_{\rm MF}(\nu) \over 2} {d\ln \sigma_M \over dM}, \cr\cr
 {\partial^n P(M,\delta_c) \over \partial \delta_c^n} &=& {(-1)^n (n-1)! \over 2 \delta_c^n}\sum^{n-1}_{m=0}
 {(-1)^m \over m!} \nu^m f_{\rm MF}^{(m)}(\nu), 
 \end{eqnarray}
 where $f_{\rm MF}(\nu)$ with $\nu = \delta_c / \sigma_M$ is a multiplicity function.
 Then,
 using the above expressions,
 the renormalized bias functions are reduced to
 \begin{eqnarray}
 c_n^\rL(k_1,\cdots,k_2) &=& {(n-1)! \over \delta_c^n f_{\rm MF}(\nu)} 
 {d \over d \ln \sigma_M}
 \left[ \left(\sum^{n-1}_{m=0} {(-1)^m  \over m! } \nu^m f_{\rm MF}^{(m)} (\nu) \right)
 W(k_1R) \cdots W(k_n R)
 \right] \cr\cr
 &=& b_n^\rL(M) W(k_1R)\cdots W(k_2R) \cr\cr
 &&+  {(n-1)! \over \delta_c^n} 
 \left(\sum^{n-1}_{m=0} {\delta_c^m \over m! } b_m^\rL(M) \right)
 {d \over d\ln \sigma_M} \left[ 
 W(k_1R) \cdots W(k_2R)
 \right],
 \label{eq:renormalizedbias}
 \end{eqnarray}
 where the scale independent function, $b_n^\rL(M)$, given by
 \begin{eqnarray}
 b_n^\rL(M) = \left( {-1 \over \sigma_M}\right)^n {f_{\rm MF}^{(n)}(\nu) \over f_{\rm MF}(\nu)}.
 \end{eqnarray}

We can take into account the effect of the primordial non-Gaussianity
on the renormalized bias function by using the non-Gaussian multiplicity function.
We discuss the non-Gaussian corrections to the renormalized bias function
in a later section.
For example,
we have 
\begin{eqnarray}
f_{\rm MF}(\nu) = f_{\rm PS}(\nu) = \sqrt{2 \over \pi} \nu e^{-\nu^2 / 2},
\end{eqnarray}
in the original PS theory, and
\begin{eqnarray}
f_{\rm MF}(\nu) = f_{\rm ST}(\nu) = A(p) \sqrt{2 \over \pi} \left[ 
1 + {1 \over (q\nu^2)^p}\right] \sqrt{q} \nu e^{-q \nu^2 / 2},
\end{eqnarray}
for the Sheth-Tormen (ST) fitting formula with $p=0.3$, $q=0.707$ and $A(p) = [1+\pi^{-1/2}2^{-p} \Gamma(1/2-p)]^{-1}$.

\section{Scale-dependent bias with primordial non-Gaussianities}
\label{sec:nGeffect}

Here, by using the simple expressions for the multipoint propagators on
large scales, which are given in the previous section,
let us write down the bias parameter with respect to the non-linearity parameters $f_{\rm NL}$,
$g_{\rm NL}$ and $\tau_{\rm NL}$.
Defining a bias parameter as
\begin{eqnarray}
P_X(k) \equiv b_X^2(k) P_\rL(k),
\end{eqnarray}
on large scales ($k \to 0$),
we have
\begin{eqnarray}
b_X^2(k) & \approx & b_{1}(M)^2 + \Delta b_X^2(k) \cr\cr
\Delta b_X^2(k)
& \equiv&
4 f_{\rm NL}  {b_{1}(M) \over {\cal M}(k)} \int {d^3 p_1 \over (2\pi)^3}
c_{2}^\rL(-\bp_1,  \bp_1) P_\rL(p_1)\cr\cr
&&+ \left( 6 g_{\rm NL} + {50 \over 9}\tau_{\rm NL} \right)  {b_{1}(M) \over {\cal M}(k)}
\int {d^3 p_1d^3 p_2 \over (2\pi)^{6}} \cr\cr
&& \qquad \times
c_{3}^\rL(-\bp_1, -\bp_2, \bp_1+\bp_2){\cal M}(p_1){\cal M}(p_2){\cal M}(|\bp_1+\bp_2|) P_\Phi(p_1)P_\Phi(p_2)\cr\cr
 && + {25 \over 9} \tau_{\rm NL} {1 \over {\cal M}(k)^2}\int {d^3 p_1 d^3 p_2 \over (2\pi)^6 }
 c_{2}^\rL(\bp_1, - \bp_1) c_{2}^\rL(\bp_2, - \bp_2)
P_\rL(p_1) P_\rL(p_2).
 \label{eq:fullpower}
\end{eqnarray}
where
$b_{1}(M)\equiv 1+c_{1}^\rL$ is the scale-independent Eulerian linear bias parameter
and $\Delta b_X^2(k)$ denotes a scale-dependent part due to the primordial non-Gaussianity.
The renormalized bias functions up to the third order are given by
\begin{eqnarray}
c_2^\rL (-\bp_1,\bp_1) &=& b_2^\rL(M) W(p_1R)^2 + 2{ 1 + \delta_c\left(b_1(M)-1\right) \over \delta_c^2}
W(p_1R){dW(p_1R)  \over d\ln \sigma_M}, \cr\cr
c_3^\rL(-\bp_1,-\bp_2, \bp_1 + \bp_2 ) &=& b_3^\rL(M) W(p_1R)W(p_2R)
W(|\bp_1 + \bp_2|R) \cr\cr
&&+ {2 + 2 \delta_c \left( b_1(M) - 1 \right) + \delta_c^2 b_2^\rL(M) \over \delta_c^3}
{d \over d\ln \sigma_M} \left[ W(p_1R)W(p_2R)
W(|\bp_1 + \bp_2|R) \right].
\end{eqnarray}
Substituting these expressions
into Eq. (\ref{eq:fullpower}), we have
\begin{eqnarray}
\Delta b_X^2(k) & \approx & 4 f_{\rm NL}  { b_{1}(M) \over {\cal M}(k)}
\left[ b_2^\rL(M) + 2 { 1 + \delta_c (b_1(M) - 1) \over \delta_c^2 } \right] \sigma_M^2
\cr\cr
&& + \left( g_{\rm NL} + {25 \over 27}\tau_{\rm NL}\right){ b_1(M)  \over {\cal M}(k)} \cr\cr
&& \qquad \times
\left[ b_3^\rL (M) +  {2 + 2 \delta_c (b_1(M)-1) +\delta_c^2 b_2^\rL(M) \over \delta_c^3}
\left( 4 + {d \ln S_3(M) \over \ln \sigma_M} \right)
 \right] \sigma_M^4 S_3(M) \cr\cr
&&
+ {25 \over 9}\tau_{\rm NL}{1 \over {\cal M}(k)^2}
\left[ b_2^\rL(M) + 2 { 1 + \delta_c (b_1(M) - 1) \over \delta_c^2 } \right]^2 \sigma_M^4,
\label{eq:simpleform}
\end{eqnarray}
where $\sigma_M$ is a variance of the smoothed density field, $\delta_M$, and
$S_3(M)$ is a skewness given by
\begin{eqnarray}
S_3(M) = {6 \over \sigma_M^4}
\int {d^3 p_1 d^3 p_2 \over (2 \pi)^6} W(p_1R)W(p_2R)W (|\bp_1 + \bp_2| R)
{\cal M}(p_1){\cal M}(p_2) {\cal M}(|\bp_1 + \bp_2|) P_\Phi(p_1) P_\Phi(p_2).
\end{eqnarray}
%
The first line in the above expression
has been already shown in Ref.~\cite{Matsubara:2012nc}
and gives the scale-dependence with $\Delta b(k) \propto 1/k^2$.
A recent paper by Biagetti, Desjacques and Riotto (2012) \cite{Biagetti:2012xy}
have shown a non-Gaussian correction to the bias including the 
effect of $\tau_{\rm NL}$ and
we find that
their expression of the non-Gaussian correction coincides with
the second line in our Eq.~(\ref{eq:simpleform}), by employing the PS mass function. 
Gong and Yokoyama (2011) \cite{Gong:2011gx} derived the similar correction term
dependent on $\tau_{\rm NL}$ as the last term in Eq. (\ref{eq:simpleform})
by making use of the high peak limit formalism.
By employing PS mass function and taking high peak limit,
we find that the last term in our expression would agree with the $\tau_{\rm NL}$-correction
term
in Ref. \cite{Gong:2011gx}.
In this sense that the results in the previous works are included in
Eq. (\ref{eq:simpleform}) as limiting expressions,
this expression is a general expression of the scale-dependent bias
depending on the non-linearity parameters, $f_{\rm NL}$, $g_{\rm NL}$ and $\tau_{\rm NL}$.

\subsection{$f_{\rm NL}$ vs $g_{\rm NL}$}

\begin{figure}[htbp]
  \begin{center}
    \includegraphics{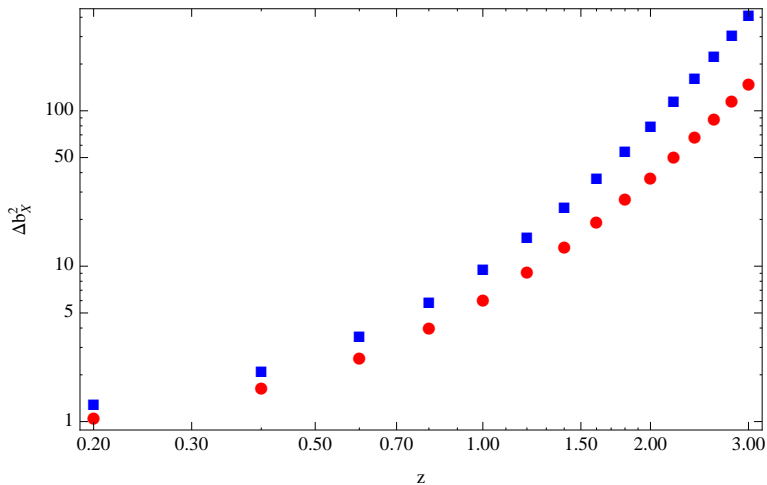}
  \end{center}
  \caption{$\Delta b_X^2 $ as functions of the redshift with
  fixing $k  = 0.005 ~h ~{\rm Mpc}^{-1}$ and $M = 5 \times 10^{13}~h^{-1} M_{\odot}$.
  We set $f_{\rm NL}=40$ for red circles and $g_{\rm NL} = 5 \times 10^{5}$ for blue boxes.
}
 \label{fig:red_dep_power.eps}
\end{figure}

Let us focus on $g_{\rm NL}$-correction term.
From Eq. (\ref{eq:simpleform}), we find
the $g_{\rm NL}$-correction gives the same scale-dependence as the
$f_{\rm NL}$-correction.
In this sense, the non-linearity parameters $g_{\rm NL}$ and $f_{\rm NL}$
are degenerate with each other in the observation of the biased objects.
One way to distinguish these two non-linearity parameters
is to see the mass-dependence ($b_1(M)$-dependence) of the non-Gaussian corrections,
which has been shown in Ref. \cite{Nishimichi:2012da}.
Here, we focus on the redshift-dependence at fixed scale of
the non-Gaussian corrections.
In Fig. \ref{fig:red_dep_power.eps},
we plot $\Delta b $ as functions of the redshift with
  fixing $k  = 0.01~h~{\rm Mpc}^{-1}$ and $M = 5 \times 10^{13}~h^{-1} M_{\odot}$.
  We set $f_{\rm NL}=40$ for red circles and $g_{\rm NL} = 5 \times 10^{5}$ for blue boxes.
%
%
%
From this figure, we find that the redshift dependence of the scale-dependent part of the bias
parameter
is different between the cases with non-zero $f_{\rm NL}$ and $g_{\rm NL}$,
and also we can see that the higher redshift observations would be a powerful tool to obtain
a constraint for $g_{\rm NL}$.

\subsection{$f_{\rm NL}$ vs $\tau_{\rm NL}$}

Next, let us focus on the $\tau_{\rm NL}$-correction terms.
As we have mentioned, a significant constraint on the $\tau_{\rm NL}$ parameter
is a powerful probe of the origins of the primordial adiabatic fluctuations,
since it has a special inequality with $f_{\rm NL}$.
To focus on the inequality between $f_{\rm NL}$ and $\tau_{\rm NL}$ in the context of the scale-dependent bias, we introduce
the power spectrum of the matter density field, $P_\rM$, and also the cross power spectrum, $P_{\rM X}$.
On large scales where the non-linear evolution of the matter density field is negligible,
 these power spectra are approximately
written as
\begin{eqnarray}
P_\rM(k) &\approx& P_\rL(k), \cr\cr
P_{\rM X}(k) & \approx & b_{1}(M) P_\rL(k)+ 2 f_{\rm NL}  {P_\rL(k) \over {\cal M}(k)} \int {d^3 p_1 \over (2\pi)^3}
c_{2}^\rL(-\bp_1,  \bp_1) P_\rL(p_1)\cr\cr
&&+ \left( 3 g_{\rm NL} + {25 \over 9}\tau_{\rm NL} \right) {P_\rL(k) \over {\cal M}(k)}
\int {d^3 p_1d^3 p_2 \over (2\pi)^{6}} \cr\cr
&& \qquad \times
c_{3}^\rL(-\bp_1, -\bp_2, \bp_1+\bp_2){\cal M}(p_1){\cal M}(p_2){\cal M}(|\bp_1+\bp_2|) P_\Phi(p_1)P_\Phi(p_2).
\end{eqnarray}
%
\begin{figure}[htbp]
  \begin{center}
    \includegraphics{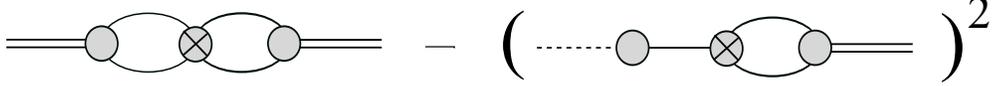}
  \end{center}
  \caption{The diagrammatic picture for the stochasticity parameter $\tilde{r}(k)-1$. The double solid line, the dashed line, the solid line and the circle
represent respectively $\delta_X$, $\delta_\rM$, $\delta_\rL$ and the multipoint propagator. 
}
 \label{fig:stochasdiagram.eps}
\end{figure}
\begin{figure}[htbp]
  \begin{center}
    \includegraphics{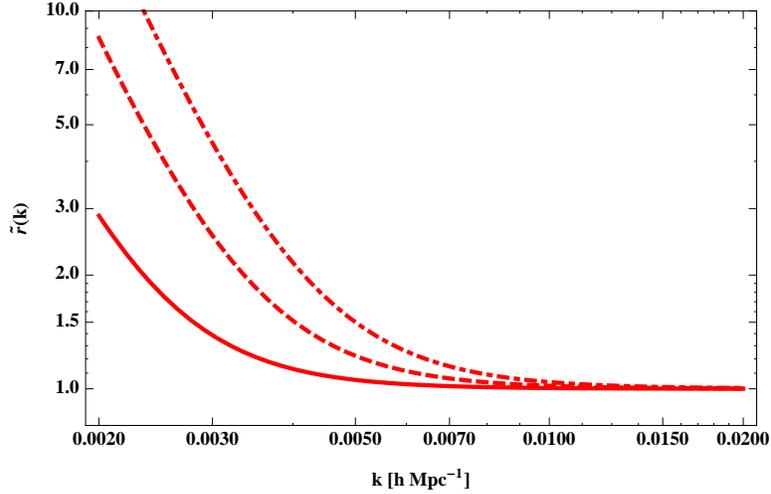}
  \end{center}
  \caption{The stochasticity parameter
as a function of $k$ with fixing $z=1$, $M = 5 \times 10^{13}~ h^{-1} M_\odot$
and $f_{\rm NL}=40$. 
The red thick line is for the case with $\tau_{\rm NL} = 2 \times 36 f_{\rm NL}^2 / 25$. 
The red dashed line is for 
$\tau_{\rm NL} = 5 \times 36 f_{\rm NL}^2 / 25$ and the red dotted-dashed line for $\tau_{\rm NL} = 10 \times 36 f_{\rm NL}^2 / 25$. 
}
 \label{fig:stochasticity_tnl.eps}
\end{figure}
By making use of the matter power spectrum and
the cross power spectrum, 
we estimate a stochasticity~(see, e.g., \cite{Matsubara:2011ck,Matsubara:2007wj,Dekel:1998eq}).
Here, for convenience, we define a stochasticity parameter as
%
\footnote{
In Refs. \cite{Matsubara:2011ck,Matsubara:2007wj,Dekel:1998eq},
the stochasticity parameter is defined as $r(k) \equiv P_{\rM X}(k) / \sqrt{P_\rM(k)P_X(k)}$.
In Ref. \cite{Smith:2010gx}, the authors introduce 
$r(k) \equiv P_X(k)/P_\rM(k) - (P_{\rM X}(k) / P_\rM(k))^2$
and Ref. \cite{Tseliakhovich:2010kf} gives $\chi(k) \equiv P_{\rM X}(k)^2 / (P_X(k) P_\rM(k))$.
Note that 
these parameters have the similar dependence on the non-linearity parameters
although each definition is slightly different.
}
\begin{eqnarray}
\tilde{r}(k) \equiv {P_\rM(k) P_X(k) \over P_{\rM X}(k)^2}.
\end{eqnarray}
At the leading order, we can obtain
\begin{eqnarray}
\tilde{r}(k)   \simeq 1 +  \left( {25 \over 9} \tau_{\rm NL} - 4 f_{\rm NL}^2 \right)
{1 \over b_1(k)^2 {\cal M}(k)^2} 
\left[ \int {d^3 p \over (2 \pi)^3} c_2^\rL(\bp, - \bp) P_\rL(p) \right]^2.
\label{eq:sto}
\end{eqnarray}
The diagrammatic picture of the stochasticity parameter $\tilde{r}(k)-1$ is shown in Fig. \ref{fig:stochasdiagram.eps}, following the rules given in Ref. \cite{Matsubara:2011ck}.
In this figure, the double solid line, the dashed line, the solid line and the circle
represent respectively $\delta_X$, $\delta_\rM$, $\delta_\rL$ and the multipoint propagator. 
In Fig. \ref{fig:stochasticity_tnl.eps}, we plot the stochasticity parameter
as a function of the wavenumber, $k$, with fixing $z=1$, $M = 5 \times 10^{13} ~h^{-1} M_\odot$
and $f_{\rm NL}=40$. The red thick line is for the case with $\tau_{\rm NL} = 2 \times 36 f_{\rm NL}^2 / 25$. 
The red dashed line is for 
$\tau_{\rm NL} = 5 \times 36 f_{\rm NL}^2 / 25$ and the red dotted-dashed line for $\tau_{\rm NL} = 10 \times 36 f_{\rm NL}^2 / 25$. 
From this figure, we find that due to the deviation from the equality as $\tau_{\rm NL}
= 36f_{\rm NL}^2/25$ the stochasticity parameter $\tilde{r}(k)$ deviates from unity on large scales.
This means that the observations of the stochasticity parameter on large scales
could be a powerful tool to study the inequality between $f_{\rm NL}$ and $\tau_{\rm NL}$.
This fact was also discussed in Ref. \cite{Smith:2010gx,Tseliakhovich:2010kf} for two-field inflation model with different parameterization.
Note that the stochasticity includes various uncertainties due to the
higher order corrections~\cite{Matsubara:2011ck} and also astrophysical processes.
The above discussion about the stochasticity given by Eq. (\ref{eq:sto})
is
%
ideal
 and hence the detection of the effect of $\tau_{\rm NL}$
needs more careful investigation.
In later section, we discuss the higher order corrections into the stochasticity.

\section{Higher order corrections}
\label{sec:correction}

In the context of the iPT,
the correction terms from the non-zero $g_{\rm NL}$ and $\tau_{\rm NL}$ discussed
in the previous section are two-loop order corrections,
while the $f_{\rm NL}$-correction term is the one-loop order.
Of course,
we have other higher order corrections which we have neglected in the previous section.
Although it is difficult to check all the contributions analytically,
in the large scale limit where the non-linear evolution of the
matter density field is negligible,
the estimation for such contributions becomes somewhat easier.
Here, as the higher order corrections,
we focus on the corrections in the pure Gaussian fluctuations
and also those
 which are linearly proportional to
the non-linearity parameters and discuss in what case we do not need carefully to treat
such higher order corrections. 

\subsection{Non-Gaussian corrections in multipoint propagators}

As we have discussed,
the multipoint propagators on large scales are rewritten in terms of the renormalized bias function as
Eq. (\ref{eq:multiren})
and
the renormalized bias functions are given by
Eq. (\ref{eq:renormalizedbias}) in terms of the multiplicity function, $f_{\rm MF}(\nu)$.
In case the primordial non-Gaussianity exists, it has been known that
the multiplicity function is modified (see, e.g., Ref. \cite{LoVerde:2011iz})
and hence the multipoint propagators should be also dependent on the non-linearity parameters.
To include the corrections due to the primordial non-Gaussianity,
we introduce a correction function as $R^{\rm NG}$ and
rewrite
the multiplicity functions as $f_{\rm MF} = f_{\rm PS (ST)} \times R^{\rm NG}$.
There are several works about $R^{\rm NG}$ analytically and numerically.
For example, 
based on the Edgeworth expansion formula, 
up to the terms which are linearly proportional to the non-linearity parameters,
we have
\begin{eqnarray}
R^{\rm NG}_{\rm Ed} \simeq 1 + \left( 
{\kappa_3(M) \over 6}H_3(\nu)  + {1 \over \nu} { \partial \kappa_3(M) / \partial M \over 6\partial \ln \sigma_M / \partial M} H_2(\nu)
\right)+ \left( 
{\kappa_4(M) \over 24} H_4(\nu) +{1 \over \nu} { \partial \kappa_4(M) / \partial M \over 24\partial \ln \sigma_M / \partial M} H_3(\nu)
\right),
\end{eqnarray}
with $\kappa_n = \langle \delta_R^n\rangle_{\rm c} / \sigma_M^{n}$ and $H_n(x)$ being
Hermite polynominal.
The higher order moments
$\kappa_3$ and $\kappa_4$ are respectively proportional to the non-linearity parameters $f_{\rm NL}$ and $g_{\rm NL}$ (also $\tau_{\rm NL}$).
As another example,
Matarrese, Verde and Jimenez (2000) \cite{Matarrese:2000iz}
has shown a fitting formula for $R_{\rm NL}$ which is given by
\begin{eqnarray}
R^{\rm NG}_{\rm MVJ}  = \left[ 
\delta_3 -  {\nu \over \delta_3}{\kappa_3(M) \over 6}  - { \partial \kappa_3(M) / \partial M \over 6
\partial \ln \sigma_M / \partial M  }  
\right] 
\left[ 
\delta_4 -  {\nu^2 \over \delta_4}{\kappa_4(M) \over 12} - { \partial \kappa_4(M) / \partial M \over 24
\partial \ln \sigma_M / \partial M  }  
\right] 
\exp\left[ {\kappa_3(M) \over 6}\nu^3+{\kappa_4(M) \over 24}\nu^4\right] 
,
\end{eqnarray}
with $\delta_n = \sqrt{1 - 2 \nu^{n-2} \kappa_n(M) / n!}$.

\begin{figure}[htbp]
  \begin{center}
    \includegraphics{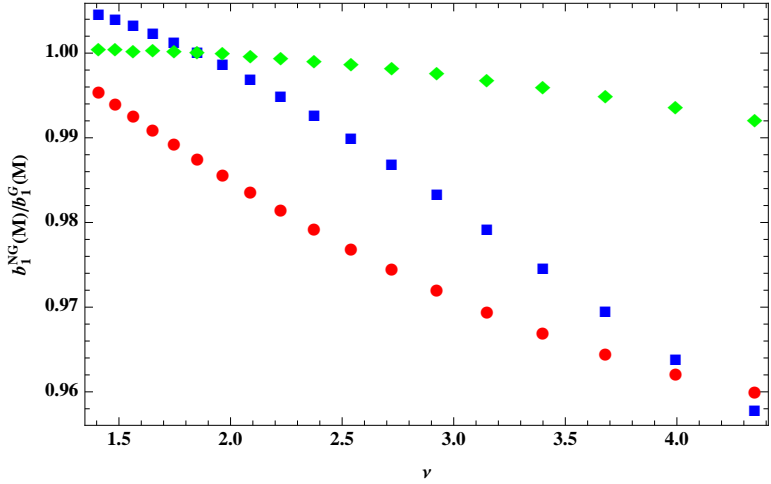}
  \end{center}
  \caption{The 
  ratio between the non-Gaussian first order bias parameter and the Gaussian one,
  $b_{1}^{\rm NG}(M) / b_{1}^{\rm G}(M)$ as a function of $\nu$. The red circles are for the $f_{\rm NL}=40$ case, the blue boxes for the $g_{\rm NL} = 5 \times 10^5$ case
  and the green diamonds for the $\tau_{\rm NL} = 10 \times (6 \times 40 / 5)^2$.}
 \label{fig:ren_bias.eps}
\end{figure}

Let us evaluate the amplitudes of the modifications of the renormalized bias functions due to
the primordial non-Gaussianity.
Hereinafter, we adopt the Sheth-Tormen fitting formula as a Gaussian mass function
and the Edgeworth expansion formula as a non-Gaussian correction of the mass function.
In this case,
the first order renormalized bias function on large scales ($k \to 0$) is
\begin{eqnarray}
c_{1,{\rm NG}}^\rL(k)& \approx& b_{1,{\rm NG}}^\rL(M) = \left({-1 \over \sigma_M} \right) {f_{\rm MF}'(\nu)\over f_{\rm MF}(\nu)}
=  \left({-1 \over \sigma_M} \right) \left[ 
{f_{\rm PS}'(\nu) \over f_{\rm PS}(\nu)} + {{R_{\rm Ed}^{\rm NG}}'(\nu) \over R_{\rm Ed}^{\rm NG}(\nu)}
\right] \cr\cr
&\simeq& 
{1 \over \delta_c}
\biggl[ q\nu^2 - 1 + {2 p \over 1 + (q\nu^2)^p} \cr\cr
&& -
{\kappa_3(M) \over 2} \left( \nu^3 -  \nu \right)
-{\kappa_4(M) \over 6} \left( \nu^4 - 3 \nu^2 \right) - {1 \over 6}
{\partial \kappa_3(M) \over \partial \ln \sigma_M} \left( 
\nu + {1 \over \nu}
\right)
- {1 \over 12} {\partial \kappa_4(M) \over \partial \ln \sigma_M} \nu^2
\biggr].
\end{eqnarray} 
In Fig. \ref{fig:ren_bias.eps}, we plot the 
ratio of the 
bias parameter in Eulerian space $b_1(M) = 1 + c_1^\rL$
with the primordial non-Gaussianity to that in the Gaussian case, as a function of $\nu$.
The red circles are for the $f_{\rm NL}=40$ case, the blue boxes for the $g_{\rm NL} = 5 \times 10^5$ case and the green diamonds for the $\tau_{\rm NL} = 10 \times (6 \times 40 / 5)^2$.
From this figure, we find that 
in the range with $ 1.5 \lesssim \nu \lesssim 4.5$
corresponding to
the mass range of the biased objects with $10^{12} \lesssim M / [h^{-1} M_\odot ] \lesssim 5 \times 10^{14} $ for $z=1$,
the contribution of the modifications of the mass function by the primordial non-Gaussianity
to the bias is negligibly small. For  $f_{\rm NL} = 40$ and $\nu=4.5$, the effect of the primordial non-Gaussianity is about $4 \%$.
For the higher peak objects ($\nu \gtrsim 5$) the effect of the primordial non-Gaussianity 
is expected to become larger and hence for much higher peak objects we could not ignore the contribution any more.
Such kind of non-Gaussian effects can be observationally seen
in the number counts of the very massive
objects which are minor components in the observations of the clustering of the biased objects. 

\subsection{Higher order loops in terms of the multipoint propagators}

In the previous subsection, 
we find that the modification of the multipoint propagators 
due to the primordial non-Gaussianity is small for 
not so high peak objects.
Hereinafter, 
we consider the case with $M = 5 \times 10^{13}~ h^{-1} M_\odot$
at $z=1.0$ and 
neglect the effects of the primordial non-Gaussianity
which appear in the multipoint propagators.

Next, let us focus on the higher order loop corrections in the power spectrum
of the biased objects, in terms of the multipoint propagators, $\Gamma_X^{(n)}$.
The power spectrum of the biased objects,
up to the terms which are linearly proportional to the non-linearity parameters,
is obtained by 
\begin{eqnarray}
P_X(k) &=& \sum_{n=1}^\infty \Biggl[ {1 \over n!}
\int {d^3 p_1 \cdots d^3 p_{n-1} \over (2\pi)^{3n}}
\Gamma_X^{(n)} (\bp_1, \bp_2, \cdots , \bk - {\bf P}_{n-1})^2
P_\rL(p_1)P_\rL(p_2) \cdots P_\rL(|\bk - {\bf P}_{n-1}|)
\cr\cr
&& +
{1 \over (n-1)!} \int {d^3 p_1 \cdots d^3 p_n \over (2 \pi)^{3n}}
\Gamma_X^{(n)} (\bp_1, \cdots, \bp_{n-1}, \bk - {\bf P}_{n-1})   \cr\cr
&& \qquad \times
\Gamma_X^{(n+1)} (-\bp_1, \cdots, -\bp_{n}, - \bk + {\bf P}_n)
P_\rL(p_1)\cdots P_\rL(p_{n-1}) B_\rL(\bk - {\bf P}_{n-1}, - \bp_n, - \bk + {\bf P}_n) \cr\cr
&&
+ {1 \over 3 (n-1)!}
\int {d^3 p_1 \cdots d^3 p_{n+1} \over (2 \pi)^{3n+3}}
\Gamma_X^{(n)} (\bp_1, \cdots, \bp_{n-1}, \bk - {\bf P}_{n-1}) \cr\cr
&& \qquad
\times
\Gamma_X^{(n+2)} (-\bp_1, \cdots,  - \bp_{n+1}, - \bk + {\bf P}_{n+1})
P_\rL(p_1)\cdots P_\rL (p_{n-1}) T_\rL (\bk - {\bf P}_{n-1}, - \bp_n, -\bp_{n+1}, - \bk + {\bf P}_{n+1})
\Biggr]
\cr\cr
&&
+ \sum_{n=2}^\infty
{1 \over 4 (n-2)!}
\int {d^3p_1 \cdots d^3p_{n-1} d^3 q \over (2 \pi)^{3n}}
\Gamma_X^{(n)}(\bp_1, \cdots, \bp_{n-1}, \bk - {\bf P}_{n-1}) \cr\cr
&& \qquad
\times
\Gamma_X^{(n)}(-\bp_1, \cdots, - \bp_{n-2}, - {\bf q} , -\bk + {\bf P}_{n-2} + {\bf q}) \cr\cr
&& \qquad\qquad\qquad\qquad\qquad
\times
P_\rL(p_1) \cdots P_\rL (p_{n-2}) T_\rL (\bp_{n-1}, \bk - {\bf P}_{n-1}, - {\bf q}, -\bk + {\bf P}_{n-2} + {\bf q}) 
 ,
 \label{eq:highpower}
\end{eqnarray}
where
${\bf P}_n = \bp_1 + \cdots + \bp_n$.
\begin{figure}[htbp]
  \begin{center}
    \includegraphics{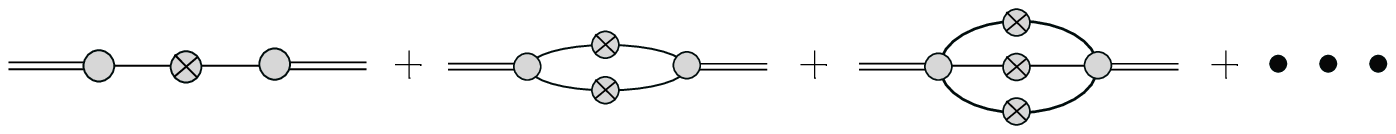}
  \end{center}
  \caption{Pure Gaussian case}
  \label{fig:gdiagram.eps}
%
  \begin{center}
    \includegraphics{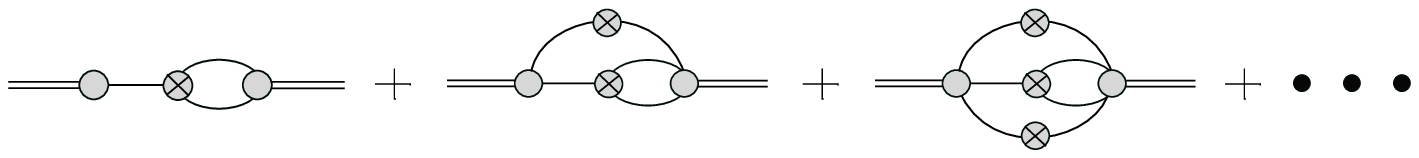}
  \end{center}
  \caption{Diagrams linearly proportional to the primordial bispectrum.}
  \label{fig:fnldiagram.eps}
\end{figure}
\begin{figure}[htbp]
  \begin{center}
    \includegraphics{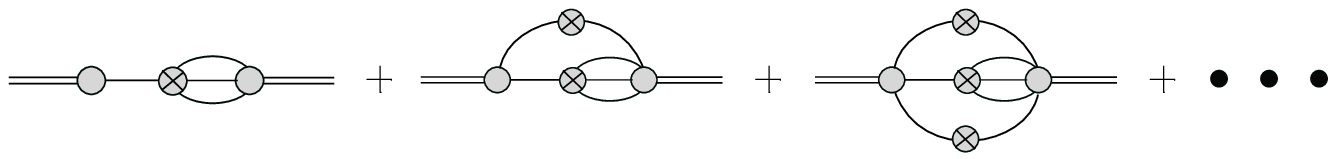}
  \end{center}
  \caption{Diagrams linearly proportional to the primordial trispectrum.}
  \label{fig:gnldiagram.eps}
  \begin{center}
    \includegraphics{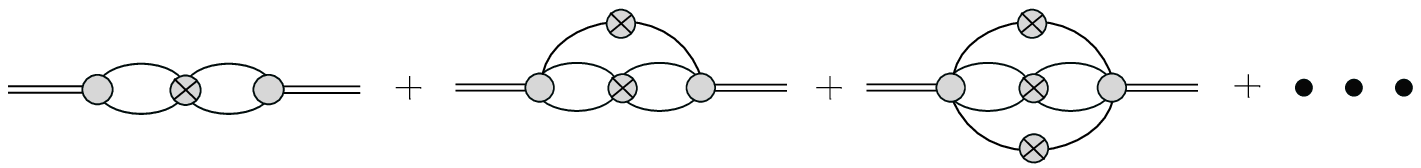}
  \end{center}
  \caption{Other diagrams linearly proportional to the primordial trispectrum.}
  \label{fig:taunldiagram.eps}
\end{figure}
In the right hand side of the above equation,
the first term in the bracket appears even in the case with pure Gaussian fluctuations, corresponding to
Fig.~\ref{fig:gdiagram.eps}.
The second term in the bracket represents the contributions which are linearly proportional to 
the primordial bispectrum parameterized by $f_{\rm NL}$.
This term is corresponding to Fig.~\ref{fig:fnldiagram.eps}.
Both the third and last terms are the contributions which are linearly proportional to
the primordial trispectrum parameterized by $g_{\rm NL}$ and $\tau_{\rm NL}$.
These terms are respectively shown by Figs. \ref{fig:gnldiagram.eps} and 
\ref{fig:taunldiagram.eps}.
Such kind of higher order corrections for the power spectrum of the matter density fields
are estimated by changing $\Gamma_X^{(n)}$ to $\Gamma_\rM^{(n)}$ in the above expression.
As we have mentioned, the non-linearity of the matter density fields
becomes negligible and $\Gamma_\rM^{(1)} \to 1$
and $\Gamma_\rM^{(n)} \to 0 $ ($n \ge 2$)
 on large scales ($k \to 0$).
Hence, 
for
the power spectrum of the matter density fields
such higher order correction terms are negligible on large scales.

\begin{figure}[htbp]
  \begin{center}
    \includegraphics{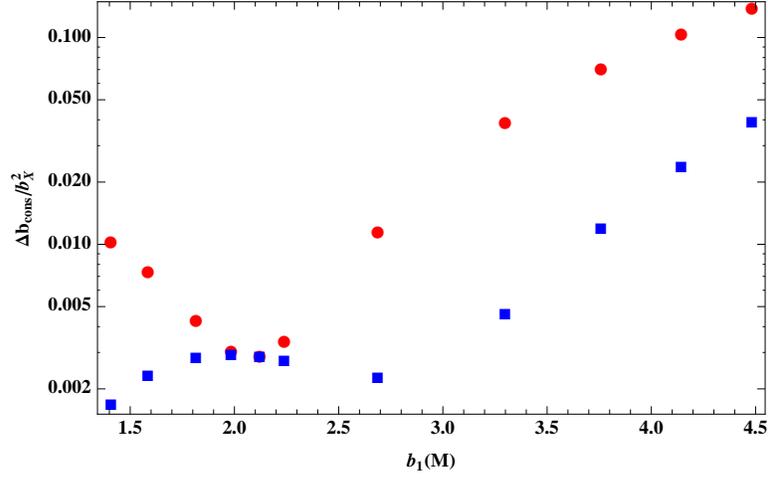}
  \end{center}
  \caption{$\Delta b_{\rm const.} / b_X^2$ as a function of the Eulerian bias $b_1(M)$ for the pure Gaussian case. We fixed $z=1.0$ and $k=0.005~h~{\rm Mpc}^{-1} $. The red circles correspond to the one-loop correction
  and the blue boxes are the two-loop correction.}
  \label{fig:highloop.eps}
\end{figure}

\begin{figure}[htbp]
  \begin{center}
    \includegraphics{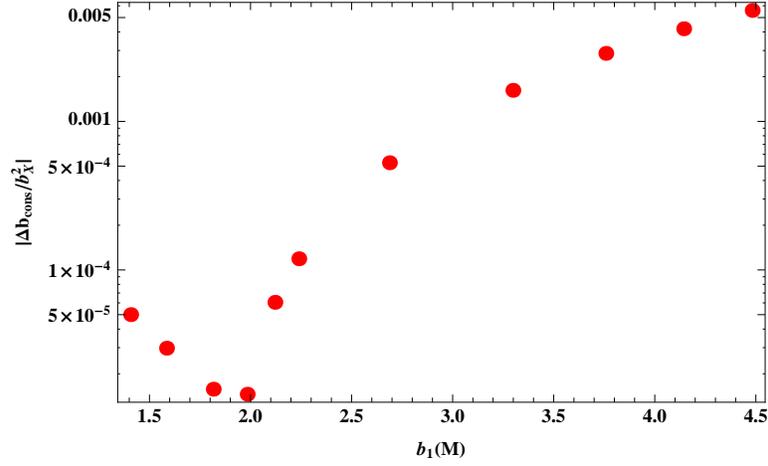}
  \end{center}
  \caption{$|\Delta b_{\rm const.} / b_X^2|$ as a function of the Eulerian bias $b_1(M)$
  with $f_{\rm NL} = 40$. We fixed $z=1.0$ and $k=0.005~h~{\rm Mpc}^{-1} $. }
  \label{fig:fnl_loop.eps}
\end{figure}

Using Eq. (\ref{eq:multiren}), the power spectrum of the biased objects including the higher order
corrections
is reduced into
\begin{eqnarray}
P_X(k) & \approx & b_{1}(k)^2 P_\rL(k)+ 4 f_{\rm NL} b_{1}(k) {P_\rL(k) \over {\cal M}(k)} \int {d^3 p_1 \over (2\pi)^3}
c_{2}^\rL(-\bp_1,  \bp_1) P_\rL(p_1)\cr\cr
&&+ \left( 6 g_{\rm NL} + {50 \over 9}\tau_{\rm NL} \right) b_{1}(k) {P_\rL(k) \over {\cal M}(k)}
\int {d^3 p_1d^3 p_2 \over (2\pi)^{6}} \cr\cr
&& \qquad \times
c_{3}^\rL(-\bp_1, -\bp_2, \bp_1+\bp_2){\cal M}(p_1){\cal M}(p_2){\cal M}(|\bp_1+\bp_2|) P_\Phi(p_1)P_\Phi(p_2)\cr\cr
 && + {25 \over 9} \tau_{\rm NL} {P_\rL(k) \over {\cal M}(k)^2}\int {d^3 p_1 d^3 q \over (2\pi)^6 }
 c_{2}^\rL(\bp_1, - \bp_1) c_{2}^\rL(\bq, - \bq)
P_\rL(p_1) P_\rL(q) \cr\cr
&&
+ P_{\rm const.}, \cr\cr
P_{\rm const.} &=&
 \sum_{n=2}^\infty
{1 \over n!} \int {d^3 p_1 \cdots d^3 p_{n-1} \over (2 \pi)^{3n-3}}
c_{n}^\rL(\bp_1,\bp_2, \cdots ,  - {\bm P}_{n-1})^2 P_\rL(p_1)P_\rL(p_2) \cdots P_\rL(| {\rm P}_{n-1}|) \cr\cr
&&
+ 2 f_{\rm NL}\sum_{n=2}^\infty 
{1 \over (n-1)!} \int{d^3 p_1 \cdots d^3 p_n \over (2\pi)^{3n}}
c_{n}^\rL (\bp_1, \cdots, \bp_{n-1}, - {\bf P}_{n-1}) \cr\cr
&& \quad \times
c_{n+1}^\rL (-\bp_1, \cdots, - \bp_n,  {\bf P}_n)
P_\rL(p_1) \cdots P_\rL(p_{n-1})
{\cal M}(|{\bf P}_{n-1}|){\cal M}(p_n){\cal M}(|{\bf P}_n|) \cr\cr
&& \qquad \times
\left[
2P_\Phi(| {\bf P}_{n-1}|)P_\Phi(p_n) + P_\Phi(| {\bf P}_{n}|)P_\Phi(p_n) 
\right]\cr\cr
&&+ 6 g_{\rm NL} \sum_{n=2}^\infty 
{1 \over 3(n-1)!} \int{d^3 p_1 \cdots d^3 p_{n+1} \over (2\pi)^{3n+3}}
c_{n}^\rL (\bp_1, \cdots, \bp_{n-1}, - {\bf P}_{n-1}) \cr\cr
&& \quad \times
c_{n+2}^\rL (-\bp_1, \cdots, - \bp_{n+1},  {\bf P}_{n+1})
P_\rL(p_1) \cdots P_\rL(p_{n-1})
{\cal M}(|{\bf P}_{n-1}|){\cal M}(p_n){\cal M}(p_{n+1}){\cal M}(|{\bf P}_{n+1}|) \cr\cr
&& \qquad \times
\left[
3P_\Phi(| {\bf P}_{n-1}|)P_\Phi(p_n)P_\Phi(p_{n+1}) + P_\Phi(p_n)P_\Phi(p_{n+1})P_\Phi(|{\bf P}_{n+1}|) 
\right] \cr\cr
&&
+ {25 \over 9}\tau_{\rm NL} \sum_{n=2}^\infty
{2 \over (n-1)!}
\int {d^3 p_1 \cdots d^3 p_{n+1} \over (2 \pi)^{3n+3}}
c_{n}^\rL(\bp_1, \cdots, \bp_{n-1},-{\bf P}_{n-1}) \cr\cr
&&
\quad \times c_{n+2}^\rL (-\bp_1, \cdots, - \bp_{n+1},{\bf P}_{n+1})
P_\rL(p_1)\cdots P_\rL(p_{n-1})
{\cal M}(|{\bf P}_{n-1}|) {\cal M}(p_n){\cal M}(p_{n+1}) {\cal M}(|{\bf P}_{n+1}|) \cr\cr
&& \qquad
\times
\left[
P_\Phi(|{\bf P}_{n-1}|)P_\Phi(p_n)P_\Phi(p_{n+1}) + P_\Phi(p_n)P_\Phi(p_{n+1})P_\Phi(|{\bf P}_{n+1}|)
\right]  \cr\cr
&& + 6 g_{\rm NL}
\int {d^3 p_1 d^3 q \over (2\pi)^6}
c_{2,{\rm G}}^\rL(\bp_1, - \bp_1) c_{2,{\rm G}}^\rL(-\bq,  \bq)
{\cal M}(p_1)^2 {\cal M}(q)^2 P_\Phi(p_1)^2 P_\Phi(q) \cr\cr
&& +
{3 g_{\rm NL} } \sum_{n=3}^\infty {1 \over (n-2)!}
\int {d^3 p_1 \cdots d^3 p_{n-1}d^3 q \over (2\pi)^{3n}}
c_{n}^\rL (\bp_1, \cdots, \bp_{n-1}, - {\bf P}_{n-1}) \cr\cr
&& \quad \times
c_{n}^\rL (-\bp_1, \cdots, -{\bf P}_{n-2}, - \bq, {\bf P}_{n-2} + \bq)
 {\cal M}(p_{n-1}) {\cal M}(|{\bf P}_{n-1}|) {\cal M}(q){\cal M} (|{\bf P}_{n-2} + \bq|)
\cr\cr
&&\qquad\times
P_\rL(p_1)\cdots P_\rL(p_{n-2}) 
\left[ 
P_\Phi(p_{n-1})P_\Phi(|{\bf P}_{n-1}|) P_\Phi(q) + P_\Phi(p_{n-1}) P_\Phi(q)P_\Phi(|{\bf P}_{n-2}+\bq|)
 \right] \cr\cr
 && +
 {25 \over 9}\tau_{\rm NL} \sum_{n=3}^\infty
 {1 \over (n-2)!} \int {d^3 p _1 \cdots d^3 p_{n-1}d^3 q \over (2\pi)^{3n}}
 c_{n}^\rL(\bp_1, \cdots, \bp_{n-1}, - {\bf P}_{n-1}) \cr\cr
 && \quad \times
 c_{n}^\rL(-\bp_1,\cdots, -\bp_{n-2}, -\bq, {\bf P}_{n-2}+ \bq) {\cal M}(p_{n-1}) {\cal M}(|{\bf P}_{n-1}|) {\cal M}(q){\cal M} (|{\bf P}_{n-2} + \bq|)
\cr\cr
&&\qquad\times
P_\rL(p_1)\cdots P_\rL(p_{n-2})  
\biggl[ P_\Phi(|{\bf P}_{n-1}|)
P_\Phi(p_{n-1})P_\Phi(q) \cr\cr
&& \qquad 
+ P_\Phi(p_{n-1})P_\Phi(|{\bf P}_{n-1}|) P_\Phi(|\bp_{n-1} - \bq|)
+ P_\Phi(q)P_\Phi(|{\bf P}_{n-1}|) P_\Phi(|\bp_{n-1} - \bq |) 
\biggr]~.
\label{eq:fullpowerhigh}
\end{eqnarray}
In the above equation, all the higher order correction terms are included in $P_{\rm const.}$
in large scale limit and 
$P_{\rm const.}$ gives the correction in the bias as
$\Delta b_{\rm const.} \equiv P_{\rm const.} / P_\rL(k) \propto 1 / P_\rL(k) \propto 1 / k$.
In the pure Gaussian case ($f_{\rm NL} = g_{\rm NL} = \tau_{\rm NL} = 0$),
the correction in the bias $\Delta b_{\rm const.} / b_X^2$ as a function of $b_1(M)$ is shown in Fig. \ref{fig:highloop.eps}.
From this figure, we find that the contributions from the higher order correction terms to the bias parameter are negligible
for the objects with not so large mass.
In case we consider here, the dimensionless power spectrum of the biased objects, which
is given by $k^3 P_X(k)$, is smaller than unity.
We could not neglect
higher order correction terms any more, when $k^3 P_X(k)$ becomes order of unity,
in which we consider the smaller scales (larger $k$) and the higher peak objects (larger $b_X^2$).
We also plot the two-loop contribution which is linearly proportional to the non-linearity parameter
$f_{\rm NL}$ in Fig. \ref{fig:fnl_loop.eps}. In this figure, we plot $|\Delta b_{\rm const.} / b_X^2|$ with $f_{\rm NL} = 40$
at $z=1.0$. We also fix the scale as $k = 0.005~h~{\rm Mpc}^{-1}$.
This figure also shows that the correction terms linearly proportional to $f_{\rm NL}$
are negligible
for the objects with not so large mass.
Here we only show the results up to the two-loop order contributions.
However, the higher order contributions than three-loops are expected to be small
for the case we consider here, because in Fig. \ref{fig:highloop.eps}
the two-loop correction is smaller than the one-loop one.
We expect that the contributions from the higher order loops
which are linearly proportional to the primordial trispectrum are also much smaller,
from the fact that the two-loop $f_{\rm NL}$-correction is smaller than the Gaussian
higher order loop corrections shown in Fig. \ref{fig:fnl_loop.eps}.
Of course, for the more precise discussion about the higher order corrections,
we need numerical calculations.

Finally, 
we discuss the contributions of the higher order correction terms 
in the stochasticity parameter. Ref. \cite{Matsubara:2011ck} have discussed
that the one-loop term could generate the stochasticity
even in the pure Gaussian case.
As we have mentioned in the previous section, the stochasticity parameter would be a powerful tool
to test the relation between $f_{\rm NL}$ and $\tau_{\rm NL}$ and hence
the higher order corrections might become "noises" for the test even on large scales.
Based on the expression for the power spectrum of the biased objects given by Eq. (\ref{eq:fullpowerhigh}),
which include the higher order corrections, 
we obtain the stochasticity parameter as
\begin{eqnarray}
\tilde{r}(k) - 1 &\simeq& \left( {25 \over 9} \tau_{\rm NL} - 4 f_{\rm NL}^2 \right)
{1 \over b_1(k)^2 {\cal M}(k)^2} 
\left[ \int {d^3 p \over (2 \pi)^3} c_2^\rL(\bp, - \bp) P_\rL(p) \right]^2 \cr\cr
&& +
{1 \over 2} {1 \over b_1(k)^2 P_\rL(k)}
\Biggl[ \int {d^3 p \over (2 \pi)^3}
c_2^\rL(\bp, - \bp)^2 P_\rL(p)^2  \cr\cr
&& \qquad
+ 
{1 \over 3} \int {d^3 p_1 \over (2 \pi)^3} {d^3 p_2 \over (2 \pi)^6}
c_3^\rL(\bp_1, \bp_2, - \bp_1 - \bp_2)^2P_\rL(p_1) P_\rL(p_2) P_\rL(|\bp_1 + \bp_2|)
\Biggr]\cr\cr
&&+ {2 f_{\rm NL} \over b_1(k)^2 P_\rL(k)}
\Biggl\{
\int {d^3 p_1 d^3 p_2 \over (2 \pi)^6}
c_2^\rL(\bp_1, - \bp_1)
c_3^\rL(\bp_1,\bp_2, - \bp_1 -\bp_2)\cr\cr
&& \qquad\qquad \times P_\rL(p_1) {P_\rL(p_2) \over {\cal M}(p_2)} 
{\cal M}(p_1){\cal M}(|\bp_1 + \bp_2|)
\left[ 2 P_\Phi(p_1) + P_\Phi(|\bp_1 + \bp_2|) \right] \cr\cr
&& 
\qquad
- {1 \over b_1(k) {\cal M}(k) }\int {d^3 p_1 \over (2 \pi)^3}
c_2^\rL(\bp_1, -\bp_1)^2 P_\rL(p_1)^2\int {d^3 p_2 \over (2 \pi)^3}
c_2^\rL (\bp_2, - \bp_2) P_\rL(p_2)
\Biggr\} \cr\cr
&& + {6 g_{\rm NL} \over b_1(k)^2 P_\rL(k)}
\int {d^3 p_1 \over (2 \pi)^3}
c_2^\rL(\bp_1, - \bp_1) P_\rL(p_1)P_\Phi(p_1) \int {d^3 p_2 \over (2 \pi)^3}
c_2^\rL(\bp_2, - \bp_2) P_\rL(p_2).
\end{eqnarray}
Here, we consider the contributions up to the two-loop order.
\begin{figure}[htbp]
  \begin{center}
    \includegraphics{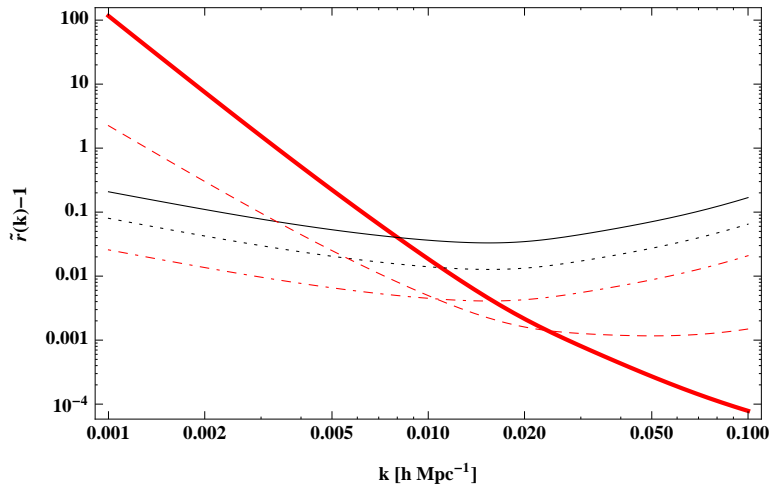}
  \end{center}
  \caption{Contribution of each term to the stochasticity parameter
as a function of the wavenumber, $k$, with fixing $z=1$ and $M = 5 \times 10^{13} ~h^{-1} M_\odot$. 
The red thick line shows the term
which is related with the inequality between the non-linearity parameters
$f_{\rm NL}$ and $\tau_{\rm NL}$ for $f_{\rm NL}=40$ and
$\tau_{\rm NL} = 5 \times 36 f_{\rm NL}^2 / 25$. The black solid and dotted lines are respectively for the one-loop and two-loop 
contributions which appear even in pure Gaussian case.
The red dashed and dotted-dashed lines are for the terms linearly proportional
to $f_{\rm NL}$, which are related with only $c_2^\rL$ (negative sign term) and $c_3^\rL$, respectively..
}
 \label{fig:stochasticity.eps}
\end{figure}
In FIg. \ref{fig:stochasticity.eps}, we plot the contribution of each term to the stochasticity parameter
as a function of the wavenumber $k$, with fixing $z=1$ and $M = 5 \times 10^{13} ~h^{-1} M_\odot$. 
The red thick line shows the term
which is related with the inequality between the non-linearity parameters
$f_{\rm NL}$ and $\tau_{\rm NL}$ for $f_{\rm NL}=40$ and
$\tau_{\rm NL} = 5 \times 36 f_{\rm NL}^2 / 25$. The black solid and dotted lines are respectively for the one-loop and two-loop 
contributions which appear even in pure Gaussian case.
The red dashed and dotted-dashed lines are for the terms linearly proportional
to $f_{\rm NL}$, which are related with only $c_2^\rL$ (negative sign term) and $c_3^\rL$, respectively.
For these parameters, we find that 
the relation between $f_{\rm NL}$ and $\tau_{\rm NL}$
would be observationally checked by large scale survey with $ k  < O(10^{-2})~h {\rm Mpc}^{-1}$.
However, as I mentioned before,
 the stochasticity includes various uncertainties due to other astrophysical processes.
The above discussion even including the higher order contributions
is still ideal and hence the detection of the effect of $\tau_{\rm NL}$
needs more careful investigation.

\section{Summary and Discussion}
\label{sec:summary}

In this paper, we have derived an accurate formula for the bias parameter
with the primordial non-Gaussianity parameterized not only by the non-linearity parameter, $f_{\rm NL}$,
but also by $g_{\rm NL}$ and $\tau_{\rm NL}$, by making use of the iPT.
The scale-dependency of the bias induced from $g_{\rm NL}$
is the same as that induced from $f_{\rm NL}$.
In this sense, it is difficult to distinguish the effect of $g_{\rm NL}$ with that of $f_{\rm NL}$.
However, we show that the redshift-dependence of the effect of $g_{\rm NL}$
is different from that of $f_{\rm NL}$
and  also
to obtain a significant constraint for $g_{\rm NL}$
the higher redshift observations would be invaluable.
Recently, as an observation of the biased object at high redshift,
an ionized fraction in the reionization era ( $6 < z < 20$)
through the 21cm observation has been proposed.
In Ref. \cite{Joudaki:2011sv,Tashiro:2012wr},  the authors shown that through such kind of the future observations
we could obtain a tight constraint for $f_{\rm NL}$.
We expect that 
such future observation can be also expected to give a tighter constraint for
$g_{\rm NL}$. 

For $\tau_{\rm NL}$, there exist a special inequality between $f_{\rm NL}$
and $\tau_{\rm NL}$.
We consider the possibility of studying this inequality through the LSS observations
and find that the stochasticity parameter deviates from unity in case
$\tau_{\rm NL} > 36 f_{\rm NL}^2 / 25$.
We expect that the future wide field survey would be a powerful tool to 
obtain a constraint for $\tau_{\rm NL}$.

In deriving the formula with the primordial non-Gaussianity parameterized by
$g_{\rm NL}$ and $\tau_{\rm NL}$,
we consider the two-loop order corrections in the context of the iPT.
In usual, these corrections can be considered as higher order effects and
neglected.
Of course, even in case with the pure Gaussian primordial fluctuation
there exist higher order contributions which may generate the extra scale-dependency
of the bias parameter.
We investigated these higher order contributions 
in the pure Gaussian case and also in the non-Gaussian case
up to the linear order in the non-linearity parameters.
We find that these higher order contributions can be negligible for the range where
the normalized power spectrum of the biased objects is smaller than unity, namely,
$k^3 P_X(k) < 1$.

Here, we just analytically derived the bias parameter with the higher order primordial
non-Gaussianity.
As a future issue, hence,
we have to check the validity of our formalism with performing numerical simulations
and it should be interesting to consider the expected constraints for $g_{\rm NL}$ and $\tau_{\rm NL}$
in the future surveys.
\\[0.5cm]

\textit{NOTE;} During the time that we were preparing this manuscript, Ref. \cite{Baumann:2012bc} appeared on the arXiv. In
Ref. \cite{Baumann:2012bc}, the authors focused on the stochasticity due to the  primordial non-Gaussianity by using the barrier crossing formalism and the peak-background split method.
They derived the stochasticity coefficient which is consistent with our result in the case with
Press-Schechter mass function.

\acknowledgments
This work was supported by the
Grant-in-Aid for JSPS Research under Grant No. 24-2775 (SY)
and also Grant-in-Aid for Scientific Research (C), 24540267, 2012 (TM).

\end{document}